\documentclass[twocolumn,showpacs,preprintnumbers,pra]{revtex4}
\usepackage{amsmath}

\usepackage{txfonts}
\usepackage{graphicx}
\usepackage{dcolumn}
\usepackage{bm}

\begin{document}

\title{Coherent destruction of tunneling in a lattice array with
controllable boundary}
\author{Liping Li$^{1}$}
\author{Xiaobing Luo$^{2}$}
\author{Xin-You L\"{u}$^{1}$}
\email{xinyoulu@gmail.com}
\author{Xiaoxue Yang$^{1}$}
\author{Ying Wu$^{1}$}
\email{yingwu2@126.com}
\affiliation{$^{1}$Wuhan National Laboratory for Optoelectronics and School of Physics,
Huazhong University of Science and Technology, Wuhan, 430074, P. R. China}
\affiliation{$^{2}$Department of Physics, Jinggangshan University, Ji'an 343009, P. R.
China}
\date{\today}

\begin{abstract}
We have investigated how the dynamics of a quantum particle
initially localized in the left boundary site under periodic driving
can be manipulated via control of the right boundary site of a
lattice array. Because of the adjustable coupling between the right
boundary site and its nearest-neighbor, we can realize either
coherent destruction of tunneling to coherent tunneling (CDT-CT)
transition or coherent tunneling to coherent destruction of
tunneling (CT-CDT) transition, by driving the right boundary site
while keeping the left boundary site driven by a periodically
oscillating field with a fixed driving parameter. In addition, we
have also revealed that our proposed CDT-CT transition is robust
against the second order coupling (SOC) between
next-nearest-neighbor sites in three-site system, whereas
localization can be significantly enhanced by SOC in four-site
system.
\end{abstract}

\maketitle

\section{introduction}

Coherent control of quantum dynamics via a periodically oscillating external
field has been one of the subjects of long-lasting interest in diverse
branches of physics and chemistry~\cite{Hanggi,Chu}. One seminal result of
the control is coherent destruction of tunneling (CDT), a phenomenon
originally discovered by Grossmann \emph{et al.} in 1991 for a periodically
driven double-well system~\cite{Grossmann}, upon the occurrence of which the
tunneling can be brought to a standstill provided that the system parameters
are carefully chosen. Due to its importance for understanding many
fundamental time-dependent processes and its potential application in
quantum motor~\cite{Salger,Hai} and quantum-information processing~\cite%
{Romero}, CDT has received growing attention from both theoretical and
experimental studies. Theoretically, it has been extended in various forms
such as nondegenerate CDT~\cite{Stockburger}, selective CDT~\cite{Villas},
nonlinear CDT~\cite{Luo}, many-body CDT~\cite{Gong,Longhi,Eckardt,Creffield}%
, instantaneous CDT~\cite{Wubs} and multiphoton CDT~\cite{Ho};
Experimentally, it has been observed in many different physical systems like
modulated optical coupler~\cite{Valle}, driven double-well potentials for
single-particle tunneling~\cite{Kierig}, three-dimensional photonic lattices~%
\cite{Zhang}, a single electron spin in diamond~\cite{Zhou}, Bose-Einstein
condensates in shaken optical lattices~\cite{Lignier, Eckardt2}, and chaotic
microcavity~\cite{Song}.

In addition to conventional approach of modulating a lattice in a uniform
fashion, modulating some certain lattice sites selectively also provides an
attractive alternative for generation of CDT, in which the rescaled
tunneling amplitudes for different sites can be shut off effectively,
thereby some intriguing quantum manipulations are realizable such as
dissipationless directed transport~\cite{Gong2} and beam splitter~\cite{Luo2}%
. CDT was traditionally thought to occur only at the isolated degeneracy
point of the quasi-energies and is related to dynamical localization\cite%
{Dunlap}. However, recently a novel quantum phenomenon called dark CDT
occurring over a wide range of system parameters has been introduced in odd-$%
N$-state systems~\cite{Luo3} which is demonstrated to be caused by localized
dark Floquet state that has zero quasi-energy and negligible population at
the intermediate state, rather than the superposition of degenerate Floquet
states. Those advances on CDT studies offer benefits for coherent control of
quantum dynamics.

In this paper, we propose a scheme for the control of quantum
tunneling in a periodically driven lattice array with controllable
boundary through combination of characteristics of both normal CDT
and dark CDT. We consider the coherent motion of a quantum particle
in a lattice array, in which the harmonic oscillating external
fields act on only the two boundary sites of chain. We have found an
interesting result that a single particle initially occupying the
left boundary site experiences coherent destruction of tunneling to
coherent tunneling (CDT-CT) transition or coherent tunneling to
coherent destruction of tunneling (CT-CDT) transition, when the
driving amplitude of the external periodic field applied to the
right boundary site is increased from zero. We have also revealed
that the CDT-CT transition in three-site lattice is robust against
the second order coupling (SOC) between next-nearest-neighbor sites,
whereas the CT-CDT transition in four-site system is significantly
affected by SOC. Moreover, we present a good understanding of the
results with help of the high-frequency approximation analysis as
well as numerical computation of the corresponding Floquet states
and quasi-energies.

\section{Model system}

\begin{figure}[htb]
\includegraphics[width=0.48\textwidth]{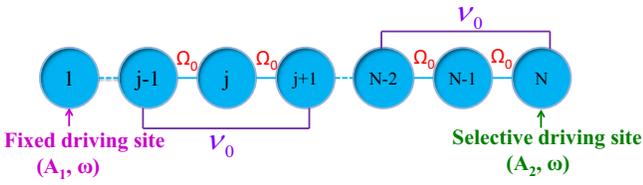}
\caption{(Color online) Schematic of a lattice array with controllable
boundary for model (\protect\ref{equ:H}). Here the left boundary site is
driven with fixed amplitude $A_1$ and frequency $\protect\omega$, and the
right boundary site driven with varied amplitude $A_2$ and fixed frequency $%
\protect\omega$. All other sites are undriven.}
\label{fig:figure1}
\end{figure}

We consider a single particle in an array of lattice sites with only
two boundary sites driven by external periodic field, as shown in
Fig.~(\ref{fig:figure1}). In our model, the left boundary site is
driven with fixed driving amplitude and driving frequency, while the
right boundary site is driven with varied driving amplitude and
fixed driving frequency. In the tight-binding approximation and
assuming a coherent dynamics, the single-particle motion can be
generally described by the tight-binding Hamiltonian

\begin{align}  \label{equ:H}
H= &E_1 (t)\left| 1 \right\rangle \left\langle 1 \right| + E_N
(t)\left| N \right\rangle \left\langle N \right| + \sum\limits_{j =
2}^{N} {\Omega _0 \left( {\left| {j - 1} \right\rangle \left\langle
j \right| + H.c.} \right)}
\nonumber \\
& + \sum\limits_{j = 2}^{N - 1} {\nu _0 \left( {\left| {j - 1} \right\rangle
\left\langle {j + 1} \right| + H.c.} \right)},
\end{align}

where $\left| j \right\rangle$ represents the Wannier state localized in the
$j$th site, $\Omega _0$ is the coupling strength connecting
nearest-neighboring sites, and $v_0$ is the second-order coupling (SOC)
strength between next-nearest-neighboring sites. Instead of modulating a
lattice in a uniform fashion, we modulate the on-site energies selectively.
In this scheme, we assume a harmonic oscillating field applied in form of $%
E_1(t)=A_1\cos (\omega t)$ for site 1, and $E_N(t)=A_2\cos (\omega t)$ for
site $N$. Here $A_1$ and $A_2$ are the driving amplitudes and $\omega$ is
the driving frequency, respectively.

We expand the quantum state of system (\ref{equ:H}) as $|\psi (t)\rangle =
\sum_{j=1}^{N} {a_j (t)\left| j \right\rangle }$, where $a_j(t)$ represents
occupation probability amplitudes at the $j$th site, with the normalization
condition $\sum\limits_j {\left| {a_j (t)} \right|^2 }=1$. From the Schr\"{o}%
dinger equation $i\partial_t|\psi (t)\rangle=H|\psi (t)\rangle$, the
evolution equation for the probability amplitudes $a_j(t)$ reads

\begin{align}  \label{equ:A}
&i\frac{{da_1 }}{{dt}} = E_1 \left( t \right)a_1 + \Omega _0 a_{2} + \nu _0
a_{3}  \nonumber \\
&i\frac{da_j}{dt} = \Omega_0 \left( a_{j -1} + a_{j+1} \right) + \nu _0
\left( a_{j-2}+ a_{j+2}\right)  \nonumber \\
&(j=2,3,...,N-1)  \nonumber \\
&i\frac{{da_N }}{{dt}} = E_N \left( t \right)a_N + \nu_0 a_{N- 2} + \Omega_0
a_{N - 1}.
\end{align}

In this work, we mainly illustrate how the quantum dynamics of a
single particle initially localized at the left boundary site can be
controlled by driving the other boundary site, for both zero and
non-zero SOC cases.

\section{CDT control by driving the right boundary site}
\subsection{CDT control in three-site system}

We start our considerations from the three-site lattice array, the minimal
one for odd-$N$-site system. According to Eq. (\ref{equ:A}), the evolution
equations for the probability amplitudes read,

\begin{align}  \label{equ:A1}
&i\frac{{da_1 }}{{dt}} = A_1\cos \left( {\omega t} \right)a_1 + \Omega_0 a_2
+ \nu_0 a_3  \nonumber \\
&i\frac{{da_2 }}{{dt}} = \Omega_0 a_1 + \Omega_0 a_3  \nonumber \\
&i\frac{{da_3 }}{{dt}} = A_2\cos \left( {\omega t} \right)a_3+\nu_0 a_1 +
\Omega_0 a_2.
\end{align}

\begin{figure}[htb]
\includegraphics[width=0.48\textwidth]{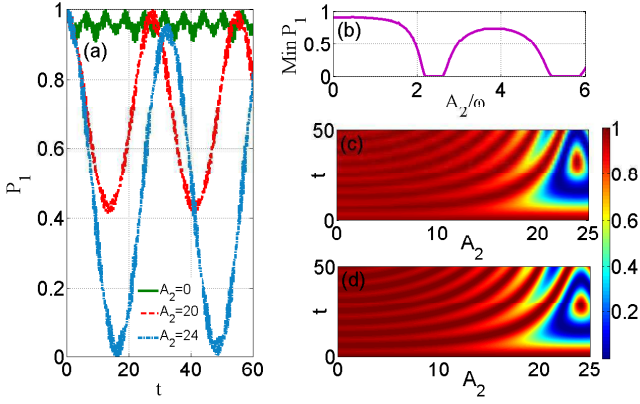}
\caption{(Color online) Three-site model (\protect\ref{equ:A1}). (a) Time
evolution of the probability at site 1, $P_1=|a_1|^2$, with different values
of $A_2$; (b) The minimum value of $P_1$ as a function of $A_2/\protect\omega
$; (c) Numerical results of probability distribution $P_1$ versus the
driving amplitude $A_2$ and time, obtained from the original model (\protect
\ref{equ:A1}); (d) Analytical results of probability distribution $P_1$
versus the driving amplitude $A_2$ and time, given by the formula (\protect
\ref{equ:P1}). The initial condition is $\{a_1 = 1, a_2 = 0, a_3 = 0\}$. The
other parameters are chosen as $A_1=22$, $\protect\omega=10$, $\Omega_0=1,
\protect\nu_0=0$.}
\label{fig:figure2}
\end{figure}

We numerically solve the time-dependent Schr\"{o}dinger equation (\ref%
{equ:A1}) with $v_0=0$, i.e., neglecting next-nearest-neighbor
tunneling in the chain. The initial state is $(1, 0, 0)^T$, and the
driving parameters of site 1 fixed as $A_1=22, \omega=10$. The
evolution of the probability distribution $P_1=|a_1|^2$ is presented
in figure (\ref{fig:figure2}) for three typical driving conditions
of site 3. For $A_2/\omega = 0$, $P_1$ remains near unity, signaling
suppression of tunneling. This is the quantum phenomenon well known
as CDT. For $A_2/\omega = 2.0$, $P_1$ oscillates
between 1 and $\sim 0.4$, showing partial suppression of tunneling. At $%
A_2/\omega$ = 2.4, $P_1$ oscillates between zero and one, demonstrating no
suppression of tunneling. The numerical results clearly indicate that the
system undergoes a CDT-CT transition when $A_2/\omega$ is increased from
zero. Such a CDT-CT transition is more clearly demonstrated in Fig.~(\ref%
{fig:figure2})(b), where the minimum value of $P_1$ is used to measure the
suppression of tunneling. When there is large suppression of tunneling, $%
\text{Min}(P_1)$ is close to 1; when there is no suppression, $\text{Min}%
(P_1)$ is zero. As clearly shown in Fig.~(\ref{fig:figure2})(b), $\text{Min}%
(P_1)$ slowly falls from its initial value to zero with $A_2/\omega$
increasing from zero to $2.4$. It is observed that the value of $\text{Min}%
(P_1)$ drops to zero in narrow intervals around the zeros of
$J_0(A_2/\omega) $, indicating that the particle is able to tunnel
freely from site to site in these regimes.

We explain the above numerical results through some analytic deduction based
on high-frequency approximation. In the high-frequency limit, we introduce
the transformation $b_1 =\exp[- i\int {A_1\cos (\omega t)}dt]a_1$, $b_2=a_2$%
, $b_3 =\exp[- i\int {A_2\cos (\omega t)}dt]a_3$, where $b_j(t) (j = 1, 2, 3)
$ are slowly varying functions. Using the expansion $\exp \left[ {\ \pm
iA\sin \left( {\omega t} \right)/\omega } \right] = \sum\nolimits_k {J_k
\left( {A/\omega } \right)\exp \left( {\ \pm ik\omega t} \right)}$ in terms
of Bessel functions and neglecting all orders except $k = 0$ in the high
frequency region, we arrive at the effective equations of motion,
\begin{align}  \label{equ:Q1}
&i\frac{{db_1 }}{{dt}} = \Omega_0 J_0 (A_1/\omega )b_2  \nonumber \\
&i\frac{{db_2 }}{{dt}} = \Omega_0 J_0 (A_1/\omega )b_1 + \Omega_0 J_0
(A_2/\omega )b_3  \nonumber \\
&i\frac{{db_3 }}{{dt}} = \Omega_0 J_0 (A_2/\omega )b_2.
\end{align}

Here we have dropped the SOC terms. It is easy to derive the analytical
solutions of Eq.~(\ref{equ:Q1}),

\begin{align}  \label{equ:Q2}
&b_1 = - \frac{{J_{02} }}{{J_{01} }}C_1 + i\frac{{J_{01} }}{{\sqrt
{J_{01}^2 + J_{02}^2 } }}\left[ {C_2 \sin \left( {Kt} \right) - C_3
\cos \left( {Kt}
\right)} \right]  \nonumber \\
&b_2 =C_2 \cos \left( {Kt} \right) + C_3 \sin \left( {Kt} \right)  \nonumber
\\
&b_3 =C_1 - i\frac{{J_{02} }}{{\sqrt {J_{01}^2 + J_{02}^2 } }}\left[
{C_2 \sin \left( {Kt} \right) - C_3 \cos \left( {Kt} \right)}
\right],
\end{align}

where $C_j(j=1,2,3)$ are constants to be determined by the initial states
and normalization condition, $J_{01}= J_0 \left( {A_1 /\omega } \right)$, $%
J_{02} = J_0 \left( {A_2 /\omega } \right)$ and $K = \Omega_0 \sqrt {%
J_{01}^2 + J_{02}^2 }$. Applying the initial conditions $b_1(0)=1$, $b_2(0)=0
$, $b_3(0)=0$ to Eq.~(\ref{equ:Q2}) yields the undetermined constants $%
C_j(j=1,2,3)$ in the forms $C_1=-\frac{{J_{01} J_{02} }}{{J_{01}^2 +
J_{02}^2 }}$, $C_2=0$, $C_3=-i\frac{{J_{01} }}{{\sqrt {J_{01}^2 + J_{02}^2 } }%
}$ and the occupation probability at site 1 as

\begin{align}  \label{equ:P1}
\left| {b_1 } \right|^2 = \left| {\frac{{J_{02}^2 }}{{J_{01}^2 + J_{02}^2 }}%
+ \frac{{J_{01}^2 }}{{J_{01}^2 + J_{02}^2 }}\cos \left( {Kt} \right)}
\right|^2.
\end{align}

From expression (\ref{equ:P1}), we immediately have two observations: (i)
when $A_2 /\omega\approx 2.4, 5.52$, i.e., the zeros of $J_{02}$, the
minimal value of $|b_1|^2$ is zero; (ii) when $A_1 /\omega$ is tuned near to
the zeros of $J_{01}$ and $A_2 /\omega$ tuned far away from those values, $%
|b_1|^2$ remains near unity. The analytical results of $P_1$ based on
formula (\ref{equ:P1}) are plotted in Fig.~(\ref{fig:figure2})(d), which
agrees well with the the numerical results obtained from the original model (%
\ref{equ:A1}) as shown in Fig.~(\ref{fig:figure2})(c).

The Floquet theory provides a powerful tool for understanding the tunneling dynamics obtained in Fig.~(\ref%
{fig:figure2}). Since Hamiltonian (\ref{equ:H}) is periodic in time,
$H(t+T)=H(t)$, where $T=2\pi/\omega$ is the period of the driving,
the Floquet theorem
allows us to write solutions of the Schr\"{o}dinger equation (\ref{equ:A}) in the form $%
a_j(t)=\tilde{a}_j(t)\exp(-i\epsilon t)$. Here $\epsilon$ is the
quasi-energy, and $\tilde{a}_j(t)$ is the Floquet state. The Floquet
states inherit the period of the Hamiltonian, and are eigenstates of
the time evolution operator over one period of the driving
\begin{align}
U(T, 0) =\mathcal{T} \exp[-i\int_0^TH(t)dt],
\end{align}
where $\mathcal{T}$ is the time-ordering operator. Noticing that
eigenvalues of $U(T, 0)$ are $\exp(-i\epsilon T)$, the
quasi-energies of this system can be numerically computed by direct
diagonalization of $U(T, 0)$.

\begin{figure}[htb]
\includegraphics[width=0.5\textwidth]{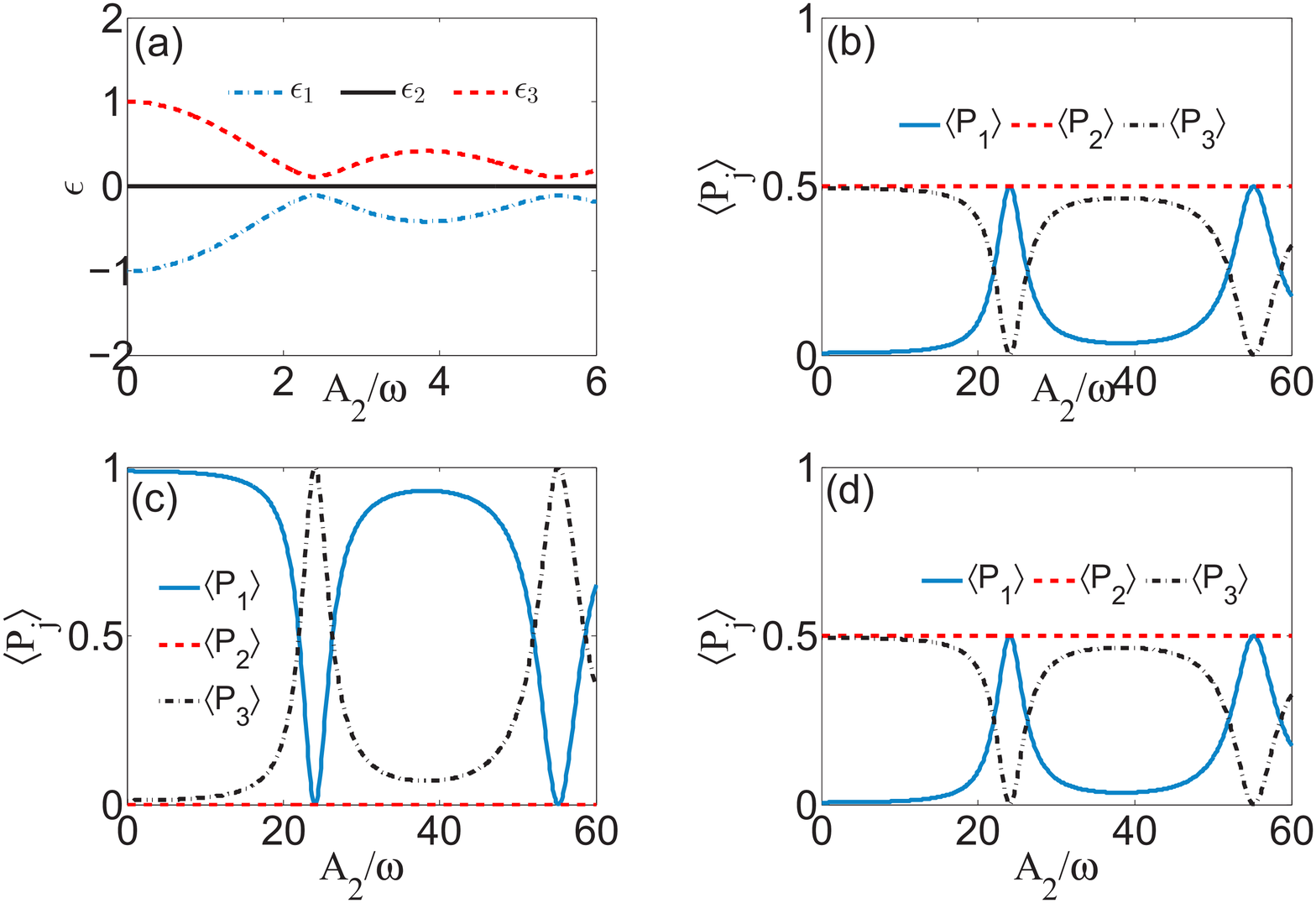}
\caption{(Color online) Quasi-energies and Floquet states of three-site
system (\protect\ref{equ:A1}). (a) Quasi-energies versus $A_2/\protect\omega$%
. The time-averaged populations for the Floquet state in the quasi-energy
level (b) $\protect\epsilon_1$, (c) $\protect\epsilon_2$ and (d) $\protect%
\epsilon_3$. The other parameters are $\Omega_0=1, \protect\nu_0=0, \protect%
\omega=10$. }
\label{fig:figure3}
\end{figure}

Our numerical results of the quasi-energies and the Floquet states
for the modulated three-site system (\ref{equ:A1}) are plotted in
Fig.~(\ref{fig:figure3}). There are three Floquet states with
quasi-energies $\epsilon_1$, $\epsilon_2$ and $\epsilon_3$. We
immediately notice from Fig.~(\ref{fig:figure3})(a) that there
exists a dark Floquet state with zero quasi-energy for all of the
values of $A_2/\omega$. For this three-site system, there is no
degeneracy in the quasi-energy levels. Therefore, the occurrence of
suppression of tunneling should be further explored through probe
into the Floquet states. We display the time-averaged population
probability $\langle P_j \rangle = (\int_0 ^T
dt|a_j|^2)/T$ for a given Floquet state $(a_1, a_2, a_3)^T$ in Figs.~(\ref%
{fig:figure3})(b)-(d). The Floquet state with $\langle P_j \rangle >0.5$ is
generally regarded as a state localized at the $j$th site. As seen in Figs.~(%
\ref{fig:figure3})(c), the dark Floquet state has negligible population at
site 2 while the population $\langle P_1 \rangle >0.5$ holds for all values
of $A_2/\omega$ except those in the vicinity of zeros of $J_0$. The other
two Floquet states are not localized at site 1 since their populations $%
\langle P_1 \rangle$ are lower than 0.5. In a word, the CDT-CT transition
shown in Fig.~(\ref{fig:figure2}) comes from the dark Floquet state, whose
population $\langle P_1\rangle$ undergoes a localization-delocalization
transition.

\subsection{CDT control in four-site system}

We are now in the position to investigate the quantum control of the
four-site system. However, similar behaviors discussed in next
subsection will appear in six-site lattice system as well. For a
four-site system, the coupled equations (\ref{equ:A}) read,

\begin{align}  \label{equ:Q3}
&i\frac{{da_1 }}{{dt}} = A_1\cos(\omega t)a_1+\Omega_0 a_2+\nu_0 a_3
\nonumber \\
&i\frac{{da_2 }}{{dt}} = \Omega_0 a_1+\Omega_0 a_3+\nu_0 a_4  \nonumber \\
&i\frac{{da_3 }}{{dt}} = \nu_0 a_1 + \Omega_0 a_2+\Omega_0 a_4  \nonumber \\
&i\frac{{da_4 }}{{dt}} = A_2\cos(\omega t)a_4+\nu_0 a_2 + \Omega_0 a_3.
\end{align}

\begin{figure}[htb]
\includegraphics[width=0.5\textwidth]{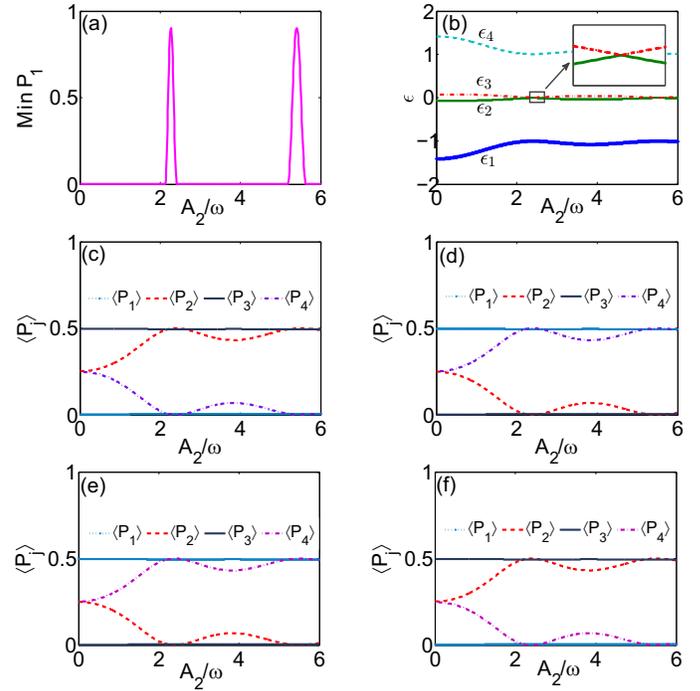} .
\caption{(Color online) Four-site model described by Eq.~(\protect\ref%
{equ:Q3}). (a) The minimum value of $P_1$ as a function of $A_2/\protect%
\omega$. The extremely narrow peaks appear at the zeros of $J_0(A_2/\protect%
\omega)$. (b) Quasi-energies versus $A_2/\protect\omega$. At the zeros of $%
J_0(A_2/\protect\omega)$, a pair of quasi-energies degenerate. (c)-(f) The
time-averaged populations $\langle P_j\rangle$ of the Floquet states
corresponding to $\protect\epsilon_1$-$\protect\epsilon_4$. The other
parameters are $A_1=22$, $\protect\omega=10$, $\Omega_0=1$, $\protect\nu_0=0$%
.}
\label{fig:figure4}
\end{figure}

We plot in Fig.~(\ref{fig:figure4})(a) the minimum value of $P_1$ versus $%
A_2/\omega$ by direct numerical simulations of the Schr\"{o}dinger equation (%
\ref{equ:Q3}) with $\nu_0=0$. We start a particle at site 1 and fix the
driving parameters of site 1 as $A_1=22, \omega=10$. As shown in Fig.~(\ref%
{fig:figure4})(a), $\text{Min}(P_1)$ takes extremely low values about zero
except at a series of very sharp peaks. The quasi-energies of this system
are shown in Fig.~(\ref{fig:figure4})(b), where we find that a pair of
quasi-energies cross at the zeros of $J_0 (A_2/\omega)$. This is the origin
of the extremely sharp peaks in localization seen in Fig.~(\ref{fig:figure4}%
)(a). Figs.~(\ref{fig:figure4})(c)-(f) show that there is no localized
Floquet states. These numerical results demonstrate the possibility of
inducing CT-CDT transition in four-site system through tuning the rescaled
driving amplitude $A_2/\omega$ from zero to the points of quasi-energy
crossing.

\subsection{Effects of second-order coupling on CDT control}

In the above discussion, the influence of second order coupling (SOC),
generally thought to be detrimental to CDT, is neglected. In this
subsection, We have checked the robustness of our proposed scheme by direct
numerical simulations of the Schr\"{o}dinger equation (\ref{equ:A}) in the
presence of SOC effects. As before, the left boundary site is initially
occupied and its driving parameters is fixed as $A_1=22, \omega=10$.
\begin{figure}[htb]
\includegraphics[width=0.5\textwidth]{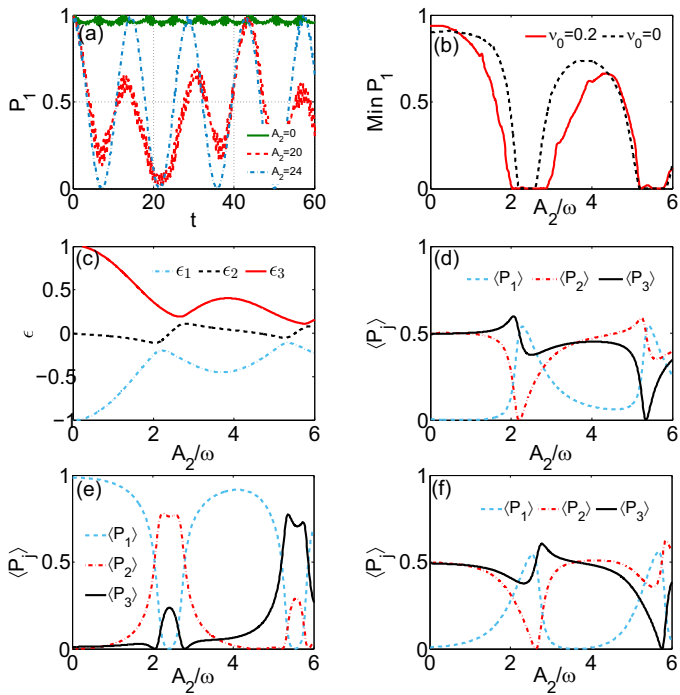} .
\caption{(Color online) Three-site model (\protect\ref{equ:A1}) with second
order coupling. (a) The time evolution of $P_1$ with different driving
amplitude $A_2$. (b) The minimum value of $P_1$ as a function of $A_2/%
\protect\omega$. (c) Quasi-energies versus $A_2/\protect\omega$. (d)-(f) The
time-averaged populations $\langle P_j\rangle$ of the Floquet states
corresponding to quasi-energies $\protect\epsilon_1$-$\protect\epsilon_3$.
The other parameters are $A_1=22$, $\protect\omega=10$, $\Omega_0=1$ and $%
\protect\nu_0=0.2$.}
\label{fig:figure5}
\end{figure}

The influence of second order coupling on three-site system is plotted in
Fig.~(\ref{fig:figure5}). When SOC effects are taken into account, the
numerical results in Figs.~(\ref{fig:figure5})(a)-(b) show that the
three-site system displays similar dynamical behaviors (CDT-CT transitions)
as the case without considering SOC. The numerically computed quasi-energies
and Floquet states are depicted in Figs.~(\ref{fig:figure5})(c)-(f).
Correspondingly, a localization-delocalization transition can be seen for
the population distribution $\langle P_1\rangle$ of the Floquet state
corresponding to a quasi-energy $\epsilon_2$ close to zero, when $A_2/\omega$
is increased from zero.

\begin{figure}[htb]
\includegraphics[width=0.5\textwidth]{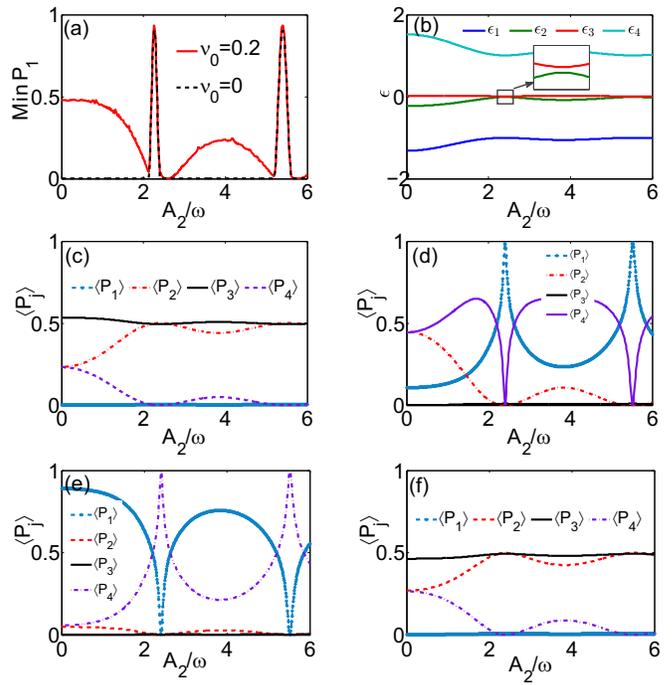}
\caption{(Color online) Four-site model (\protect\ref{equ:Q3}) with second
order coupling. (a) The minimum value of $P_1$ as a function of $A_2/\protect%
\omega$. (b) Quasi-energies versus $A_2/\protect\omega$. (c)-(f) The
time-averaged populations $\langle P_j\rangle$ of the Floquet states
corresponding to quasi-energies $\protect\epsilon_1$-$\protect\epsilon_4$.
The other parameters are $A_1=22$, $\protect\omega=10$, $\Omega_0=1$ and $%
\protect\nu_0=0.2$.}
\label{fig:figure6}
\end{figure}

We have also investigated the impact of SOC on the dynamics of four-site
system, as shown in Fig.~(\ref{fig:figure6}). Fig.~(\ref{fig:figure6})(a)
shows the minimum value of $P_1$ as a function of the driving amplitude $%
A_2/\omega$. As $A_2/\omega$ is increased from zero, $\text{Min}(P_1)$
steadily decreases from a value of $\sim 0.5$ to zero before it peaks at $%
A_2/\omega\approx 2.4$. Contrary to our expectation, SOC facilitates rather
than hindering the localization. This result is somewhat counter-intuitive.
From Fig.~(\ref{fig:figure6})(b), it can be seen that a pair of the
quasi-energies make a series of close approaches to each other as $A_2/\omega
$ increases. Detailed examination of the close approaches reveals that they
are in fact avoided crossings. At the points of close approach, the
tunneling is suppressed and the localization peaks. Compared with the zero
SOC case shown in Fig.~(\ref{fig:figure4}), the population distributions of
Floquet states change dramatically for four-site system in the presence of
SOC effects; see Figs.~(\ref{fig:figure6})(c)-(f). The modification of the
four-site-system's dynamics due to SOC, such as enhancement of localization,
is the consequence of the Floquet states localized at site 1 instead of the
quasi-energy degeneracy.

\subsection{CDT control in five- and six-site systems} Our analysis above
is given for three- and four-site systems, but similar behavior can
be obtained for other finite number of sites when the SOC effects
are not considered.

\begin{figure}[htb]
\includegraphics[width=0.5\textwidth]{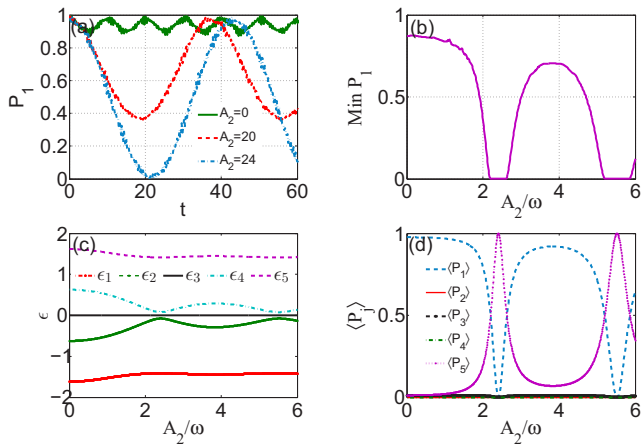} .
\caption{(Color online) Five-site model. (a) Time evolution of the
probability at site 1, $P_1=|a_1|^2$, with different values of
$A_2$; (b) The minimum value of $P_1$ as a function of
$A_2/\protect\omega$; (c) Quasi-energies versus
$A_2/\protect\omega$; (d) The time-averaged
populations $\langle P_j\rangle$ of the dark Floquet state corresponding to $%
\protect\epsilon_3$. The other parameters are $A_1=22$,
$\protect\omega=10$, $\Omega_0=1$, $\protect\nu_0=0$.}
\label{fig:figure7}
\end{figure}

\begin{figure}[htb]
\includegraphics[width=0.5\textwidth]{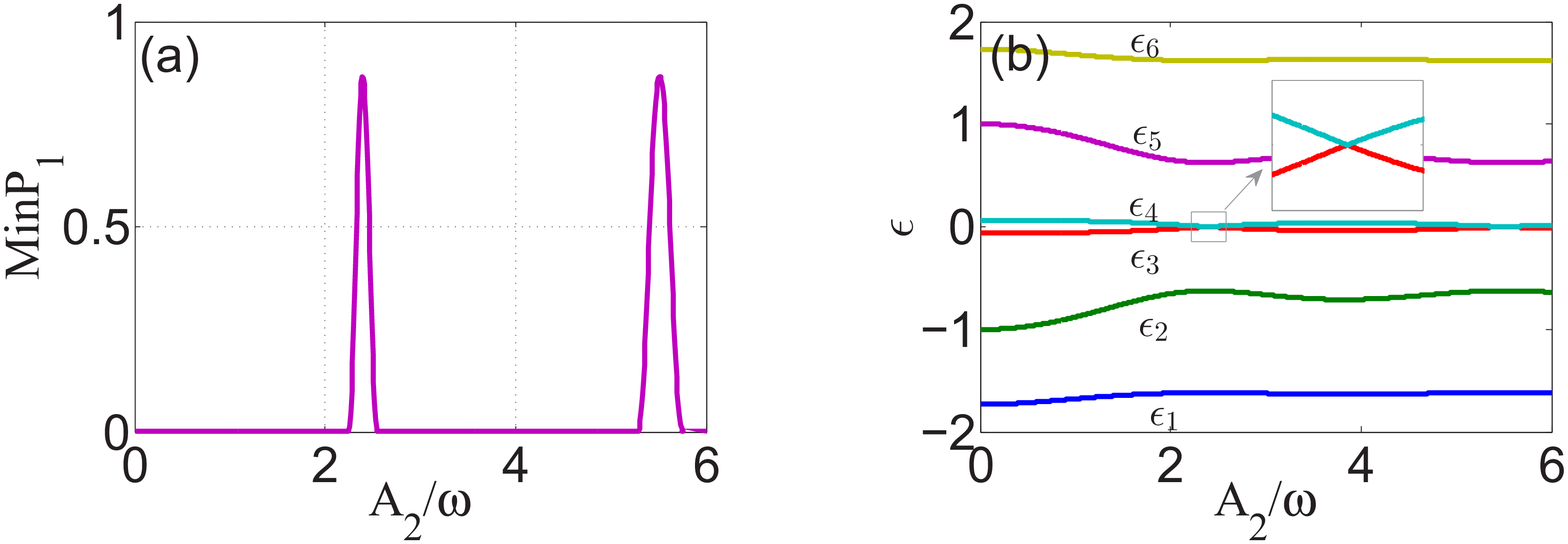} .
\caption{(Color online) Six-site model. (a) The minimum value of
$P_1$ as a
function of $A_2/\protect\omega$; (b) Quasi-energies versus $A_2/\protect%
\omega$. The other parameters are $A_1=22$, $\protect\omega=10$, $\Omega_0=1$%
, $\protect\nu_0=0$.} \label{fig:figure8}
\end{figure}

The quantum dynamics of the driven $N$-site systems is investigated
by
direct integration of the time-dependent Schr\"{o}dinger equation (\ref{equ:A})
with the particle initially localized at site 1. In what follows,
we will present the numerical results for $N=5$ and $6$.  Dynamics for five sites is depicted in Fig.~(\ref{fig:figure7}%
), which shows a similar CDT-CT transition as that of three-site
system. Like the case of three-site system, this five-site system
possesses a dark Floquet state with zero quasi-energy and negligible
population at all of the even $j$th sites, as illustrated in
Figs.~(\ref{fig:figure7}) (c)-(d). The reason for CDT-CT transition
in the five-site system lies in the fact that population
distribution $\langle P_1\rangle$ for the dark Floquet state also
experiences a localization-delocalization transition (see Fig.~(\ref%
{fig:figure7})(d)). When $N=6$, The CDT occurs only at the isolated
 points of parameters as shown in Fig.~(\ref%
{fig:figure8})(a), where a pair of quasi-energies become degeneracy
(Fig.~(\ref {fig:figure8})(b)). This is exactly the same as in
four-site system.  It is demonstrated that the CT-CDT transition can
be induced in six-site system by increasing $A_2/\omega$ from zero.

\section{Conclusion and discussion}

In conclusion, we have studied how the dynamics of a single quantum
particle initially localized in the left boundary site under
periodic driving can be controlled by only driving the right
boundary site of a lattice array. In studying the number of lattice
sites $N=3,4,5$ and $6$, we have found that (i) a dark Floquet state
with a zero quasi-energy  exists in the three- and five-site
systems. By raising the driving amplitude of the external periodic
field applied to the right boundary site from zero, we can induce a
CDT-CT transition caused by a localization-delocalization transition
of the dark Floquet state; (ii) no localized Floquet state occurs in
the four- and six-site systems. However, we can realize a CT-CDT
transition in the same way, thanks to the fact that a pair of
quasi-energies degenerate at isolate points of parameters. We have
also revealed that the CDT-CT transition in three-site lattice
persists when the SOC effects are considered. Long-range
interactions (non-nearest-neighbor couplings) in the lattice are
important and non-negligible in some systems like
biomolecules~\cite{Mingaleev}, polymer chains~\cite{Hennig}, coupled
waveguides~\cite{Luo2}, and charge transport in a quantum dot
array~\cite{Zhao, Braakman}. Therefore, our proposed scheme provides
a new route to possible application of CDT in such systems.
Nevertheless, the CT-CDT transition in four-site system is
significantly affected by SOC. In such a system, it is found, there
exists a counter-intuitive phenomenon that SOC can enhance rather
than hinder localization.

Finally, we discuss the experimental possibility of observing our
theoretical predictions. The Hamiltonians (\ref{equ:H})
can be realized in different physical systems, for example using cold atoms
or trapped ions in optical lattices~\cite{Zoller, Morsch, Schmied}, electron
transport in quantum dot chains\cite{Byrnes, Yamamoto, Braakman}, and light
propagation in an array of coupled optical waveguides with harmonic
modulation of the refractive index of the selected waveguide along the
propagation direction~\cite{Garanovich,Longhi2}. Recently, several
zigzag-type waveguide systems were experimentally realized~\cite%
{Longhi3,Szameit}, which provides simpler realization of a
one-dimensional lattice with controllable first- and second-order
(that is, next-nearest neighbors) couplings. Though experimental
observation of CDT has been reported in many different physical
systems, the proofs presented are not so rigorous as claimed,
because CDT occurs only at isolated system parameters. Since it is
impossible to precisely determine a parameter point through
experiment, what is claimed is possibly not a genuine CDT which can
be testified only after infinite time of evolution, something
unrealizable in real experiment. Due to the limited time of
evolution in experiment, the CDT as called may be a pseudo-CDT
occurring around the surrounding regime adjacent to an isolated
parameter, instead of a real one. Hence, observation of CDT, such as
the one in our considered four-site and six-site systems, still
present a generally unsolved challenge in current experimental
setups

\acknowledgments The work is supported in part by the National
Fundamental Research Program of China (Grant No. 2012CB922103), the
National Science Foundation (NSF) of China (Grant Nos. 11375067,
11275074, 11374116, 11204096 and 11405061), the Fundamental Research
Funds for the Central Universities, HUST (Grant No. 2014QN193). X.
Luo is supported by the NSF of China under Grants 11465009,
11165009, 10965001, the Program for New Century Excellent Talents in
University of Ministry of Education of China (NCET-13-0836), the
financial support from China Scholarship Council, and Scientific and
Technological Research Fund of Jiangxi Provincial Education
Department under Grant No. GJJ14566. X. Luo also would like to
acknowledge his debt to Congjun Wu for providing him with an
opportunity of visiting UCSD where part of this work is carried out.

Liping Li and Xiaobing Luo are equally contributed to this work.

\end{document}